\documentclass[11pt,preprint]{aastex}

\newcommand{\ldl}{${\lambda}/{\Delta}{\lambda}$}

\slugcomment{Accepted to ApJ Letters, v. 614}

\shorttitle{L Subdwarf 2MASS 1626+3925}
\shortauthors{Burgasser et al.}

\begin{document}

\title{Discovery of a Second L Subdwarf in the Two Micron All Sky Survey}

\author{
Adam J.\ Burgasser\altaffilmark{1}}

\affil{Department of Astrophysics,
Division of Physical Sciences,
American Museum of Natural History, Central Park West at 79$^{th}$ Street,
New York, NY 10024, USA; \email{adam@astro.ucla.edu}}

\altaffiltext{1}{Spitzer Fellow}

\begin{abstract}
I report the discovery of the second L subdwarf identified in the Two Micron All Sky Survey,
2MASS J16262034+3925190.  This high proper motion object ($\mu = 1{\farcs}27{\pm}0{\farcs}03$
yr$^{-1}$) exhibits near-infrared spectral features indicative
of a subsolar metallicity L dwarf, including strong metal hydride
and H$_2$O absorption bands, pressure-broadened alkali lines, and blue near-infrared colors
caused by enhanced collision-induced H$_2$ absorption.
This object is of later type than any of the known M subdwarfs, but does not appear to
be as cool as the apparently late-type sdL 2MASS 0532+8246.  The radial
velocity ($V_{rad} = -260{\pm}35$ km s$^{-1}$)
and estimated tangential velocity ($V_{tan} \approx$ 90-210 km s$^{-1}$)
of 2MASS 1626+3925 indicate membership in the Galactic halo,
and this source is likely near or below
the hydrogen burning minimum mass for a metal-poor star.
L subdwarfs such as 2MASS 1626+3925 are useful probes
of gas and condensate chemistry in
low-temperature stellar and brown dwarf atmospheres, but
more examples are needed to study these objects as a population
as well as to define a rigorous classification scheme.
\end{abstract}


\section{Introduction}

M dwarfs dominate the Galactic stellar population both in number and mass.
As long-lived sources, they are important tracers of Galactic star formation history
and chemical evolution, as well as the initial mass function.
Most M dwarfs identified in the vicinity of the Sun are Population I ``field'' stars,
with kinematics consistent with Galactic disk orbits and near-solar metallicities.
However, a small fraction ($\sim$ 0.3\%; Digby et al.\ 2003) are metal-poor.
These so-called subdwarfs (sd) typically exhibit halo kinematics
(${\langle}V{\rangle} = -202$ km s$^{-1}$; Gizis 1997)
and were likely formed early in the Galaxy's history.
Until recently, the latest-type subdwarfs known were M stars,
identified primarily as high proper-motion sources in multi-epoch
photographic plate surveys (e.g., Luyten 1979).
In contrast, hundreds of field dwarfs cooler than type M --- the
L dwarfs \citep{kir99,mrt99} --- have been identified
in wide-field, near-infrared (NIR) surveys such as
2MASS \citep{cut03}, DENIS \citep{epc97}, and SDSS \citep{yor00}.
The majority of these low mass stars and brown dwarfs are too faint
at optical wavelengths to have been detected on earlier photographic plates.

The first L-type subdwarf was recently identified in the 2MASS database
\citep[hereafter B03]{me03f}.  This object, 2MASS 0532+8246, exhibits many
spectral features similar to late-type L field dwarfs,
including a steep red optical slope, strong alkali lines and metal
hydride bands, deep H$_2$O absorption in the NIR,
and pressure-broadened \ion{Na}{1} and \ion{K}{1} at optical
wavelengths.  The metal hydride bands in the spectrum of 2MASS 0532+8246
are far stronger than those observed
in any field L dwarf, however.  This source also has blue NIR colors,
$J-K_s \approx 0.3$, in contrast to $J-K_s \approx 1.5-2.5$
typical of field L dwarfs \citep{kir00}.  The spectral and photometric
properties of 2MASS 0532+8246, in addition to its high
space motion, indicate that it is a L-type halo subdwarf.
A second, apparently earlier-type, L subdwarf
has also been identified by \citet{lep03b} as part of a
proper motion survey using red optical photographic plates
\citep{lep02}.  These discoveries have extended our census of metal-poor halo
objects into the substellar regime (B03).

In this Letter, I present the discovery of a second L subdwarf identified
in the 2MASS Point Source Catalog,
2MASS J16262034+3925190 (hereafter 2MASS 1626+3925).  The identification
and subsequent imaging observations of this object are described in $\S$ 2,
and its NIR spectrum is presented and analyzed in $\S$ 3.
The importance of this and future L subdwarf discoveries is addressed in $\S$ 4.

\section{Identification and Imaging Observations}

2MASS 1626+3925 was initially selected from the 2MASS database in
a search for T dwarfs, brown dwarfs cooler than spectral class L
\citep{me02a,me03a}.  Like most T dwarfs, this object has blue NIR colors
($J-K_s = -0.03{\pm}0.08$; Table 1) and no optical counterpart in the
USNO-A2.0 catalog \citep{mon98}.  However,
inspection of red ($R$) and infrared ($I_N$) photographic plates from the
Second Palomar Sky Survey \citep[hereafter POSS-II]{rei91} reveal
an offset, moving optical counterpart (Figure 1) with $R$ = 19.84
and $I_N$ = 16.68 \citep[USNO-B1.0]{mon03}.  This counterpart
was confirmed by $J$-band IRCam \citep{mur95} and
$R$-band CCD imaging observations, both conducted at the Palomar 60'' Meyer
Telescope on 14 April and 1 July 2000 (UT), respectively.  The offset between
the POSS-II $R$ plate (epoch 1993.3) and the 2MASS detection (epoch 1998.3)
implies a high proper motion,
measured\footnote{The position of 2MASS 1626+3925 on the
$R$ plate was measured by creating a 2MASS-based astrometric reference frame from
25 nearby field stars; see \citet{me04a}.  A statistically consistent
value of $\mu = 1{\farcs}37{\pm}0{\farcs}07$ yr$^{-1}$ at $\theta = 280{\pm}3\degr$
was measured between the 2MASS detection and the epoch 1995.5 POSS-II $I_N$ plate.}
to be $\mu$ = 1$\farcs$27$\pm$0$\farcs$03 yr$^{-1}$ at position angle
$\theta = 282{\fdg}6{\pm}1{\fdg}4$.
Despite being an interesting, late-type
source, 2MASS 1626+3925 was initially rejected as being too optically bright
for a T dwarf.  However, following the identification
of 2MASS 0532+8246 in the same search sample, this source was re-examined
to determine if it too is a late-type subdwarf.

\section{Near-Infrared Spectroscopy}

NIR spectroscopic observations of 2MASS 1626+3925
were obtained during 23-24 July 2004 (UT)
using the SpeX instrument \citep{ray03}
mounted on the NASA 3.0m Infrared Telescope Facility.  Observing
conditions were excellent, with clear skies and 0$\farcs$5--0$\farcs$7 seeing.
Spectral data were obtained in both prism
and cross-dispersed modes, with complete 0.7--2.5 $\micron$ coverage at spectral
resolutions {\ldl} $\approx$ 150 and 1200, respectively, for the employed 0$\farcs$5 slit.
Additional observations of the A0 V stars HD 153345 and HD 158261
were obtained for flux and telluric absorption calibration.
All spectral data were reduced using the SpeXtool package
\citep{cus04}. Further details on the experimental design and
data reduction are given in \citet{me04a}.

Figure 2 shows the reduced, low resolution NIR spectrum
of 2MASS 1626+3925, along with those of three late-type M subdwarfs---LHS 377 \citep{giz97},
LSR 2036+5059 \citep{lep03a}, and SSSPM 1013$-$3956 \citep{sch04}\footnote{We
note that SSSPM 1013$-$3956 was also identified in the same 2MASS
T dwarf search sample from which 2MASS 0532+8246 and 2MASS 1626+3925 were drawn.}---all
obtained with the SpeX spectrograph (Burgasser et al.\ in prep.).
These subdwarfs are optically classified sdM7, sdM7.5, and sdM9.5,
respectively, according to the
scheme of \citet{giz97}.  All four NIR spectra exhibit signatures typical of
late-type dwarfs, with steep red optical spectral slopes,
TiO absorption at 0.84 $\micron$,
strong alkali lines of \ion{Na}{1} and \ion{K}{1}
(particularly between 1.1 and 1.3 $\micron$), and stellar H$_2$O absorption
at 1.4 and 1.8 $\micron$.  They also exhibit features characteristic
of metal-poor stars, including enhanced metal hydride bands (FeH, CrH);
weak metal oxide bands (note the weak or absent CO band
at 2.3 $\micron$); and blue NIR spectral energy distributions.
The blue NIR colors are due to collision-induced, H$_2$ 1-0 quadrupole absorption
centered near 2.5 $\micron$, which
is enhanced in the higher-pressure photospheres
of metal-poor cool subdwarfs\footnote{The reduced metallicity of a cool subdwarf
implies reduced metal opacities and therefore a more transparent atmosphere.
The photosphere lies at a deeper, and hence higher temperature and pressure,
layer as compared to an equivalent solar-metallicity dwarf \citep{bur02}.}
\citep{sau94,bor97,leg00}.

In 2MASS 1626+3956, these low-temperature, metal-poor spectral features become
even more pronounced.  CrH and FeH bands at
0.86 and 0.87 $\micron$ are seen distinctly in the moderate resolution
cross-dispersed data, while
the 0.99 $\micron$ FeH Wing-Ford band is deeper
than those of any field M or L dwarf observed to date \citep{kir99,kir00}.
FeH absorption is also strong in the 1.2-1.3 $\micron$ region.
We detect a weak signature of the ${\Lambda}^4{\Phi}$-X$^4{\Gamma}$
band of TiH at 0.94 $\micron$ \citep[M.\ Cushing 2004, priv.\ comm.]{and03}
in the moderate resolution data, a feature that
can also be seen in the red optical spectrum of 2MASS 0532+8246
(B03, labelled as H$_2$O in their Figure 2).  This band is not responsible
for an as yet unidentified feature centered at
0.96 $\micron$, however, present in the spectra of both 2MASS 1626+3925 and 2MASS 0532+8246.
Beyond 1.3 $\micron$
the NIR spectrum of 2MASS 1626+3925 is generally featureless, with the exception
of strong H$_2$O absorption at 1.4 and possibly 1.8 $\micron$.
Indeed, the 2.3 $\micron$ CO band is completely absent,
and $K$-band flux is highly suppressed by H$_2$ absorption.

Figure 3 displays the moderate resolution 1.15-1.35 $\micron$ spectrum of 2MASS 1626+3925,
along with those of 2MASS 0532+8246 and the L2 field dwarf Kelu 1 \citep{rui97},
both measured with the Keck NIRSPEC instrument \citep[B03]{mcl03}.
Significant FeH absorption
is present in the spectra of all three objects, with
strong bands at 1.20, 1.22, and 1.24 $\micron$, and
fine features throughout the displayed spectral region \citep{cus03}.
\ion{Fe}{1} lines may also be present in the spectrum of 2MASS 1626+3925, albeit quite weak.
The \ion{K}{1} doublet lines at 1.17/1.18 and 1.24/1.25 $\micron$
are notably broadened in the subdwarf spectra.  A measurement of the full width
at half maximum of the shorter
wavelength pair yields 16.3 and 19.4 {\AA} at 1.17 and 1.18 $\micron$,
respectively, 20\% broader than
measured for the rapidly rotating Kelu 1 ($v\sin{i} = 60{\pm}5$ km s$^{-1}$;
Basri et al.\ 2000)\footnote{The same \ion{K}{1} lines are 20\% broader
in the spectrum of 2MASS 0532+8246 as compared to
its field analog, DENIS 0205$-$1159AB (B03).  The 1.24/1.25 $\micron$ \ion{K}{1}
lines are partly blended with FeH absorption.}.
This is indicative of pressure broadening, consistent with the high photospheric
pressures implied from enhanced H$_2$ absorption at $K$-band.

The spectra in Figure 2 show a natural progression of deepening FeH and
H$_2$O bands, strengthening H$_2$ absorption, and increasingly red
optical slopes.  2MASS 1626+3925 is clearly the latest-type source
of the four, apparently cooler than the sdM9.5 SSSPM 1013$-$1356.  Furthermore,
the depth of the 1.4 $\micron$ H$_2$O band is similar to those
of early- and mid-type L dwarfs observed with the same instrumental
setup \citep{me04a}, although it is somewhat weaker than that
of 2MASS 0532+8246 (Figure 3).  We therefore conclude that 2MASS 1626+3925
is a {\em bona-fide} L subdwarf, but of earlier type than 2MASS 0532+8246.

Furthermore, the kinematics of 2MASS 1626+3925 appear to be consistent with membership
in the Galactic halo.  The radial velocity of this source was measured by
cross correlating its moderate resolution $J$-band spectrum with
NIRSPEC spectra of 8 L dwarfs
with known radial velocities \citep[see B03]{bas00,rei02,mcl03,bai04}.
The resulting heliocentric value, $V_{rad} = -260{\pm}35 $km s$^{-1}$,
is much too high for a nearby disk star.  Likewise,
assuming that 2MASS 1626+3925
has an absolute $J$-band magnitude similar to that of an early- to mid-type
L dwarf, $M_J \approx 11.5-13.5$ \citep{vrb04}, its estimated spectrophotometric distance
of 15-35 pc implies $V_{tan} \approx$ 90-210 km s$^{-1}$, characteristic
of thick disk or halo stars. A more accurate determination of this object's kinematics
will require a parallax measurement.

Finally, if 2MASS 1626+3925 has the same effective temperature as an
early- or mid-type L dwarf, 1800 $\lesssim$ T$_{eff}$ $\lesssim$ 2300 K
\citep{gol04}, then it straddles the hydrogen burning minimum mass
for $Z \sim 0.1Z_{\sun}$ and is substellar for lower metallicities
(c.f., Figure 3 of B03).  Again, parallax and
bolometric luminosity measurements are needed to verify this possibility.

\section{Discussion}

The identification of L subdwarfs such as 2MASS 1626+3925 enables a new approach
to studying the atmospheric physics of cool stars and brown dwarfs.
The chemistry and bandstrengths of molecular species in
the atmospheres of late-M and L dwarfs are highly sensitive to the total
metal abundance.  Metal-poor environments can therefore produce emergent spectra
with very different dominant species (i.e., metal hydrides) or enable the
detection of species that are otherwise suppressed (i.e., TiH).
It remains unclear as to what extent condensate clouds can form in
cool metal-poor atmospheres (B03), the suppression of which would have a significant impact on
NIR spectral energy distributions and
the retention of atmospheric chemical species such as atomic alkalis \citep{lod99,bur01}.
The higher photospheric pressures of cool subdwarfs also make them excellent
laboratories for studying the pressure-broadened wings of the
\ion{K}{1} and \ion{Na}{1} fundamental transitions at optical wavelengths \citep{bur00}
as well as pressure-sensitive H$_2$ absorption in the NIR.  The latter
is a key absorber in the spectra of cool white dwarfs \citep{sau94,han98}.

The unique spectra of L subdwarfs imply that existing classification schemes are
inappropriate to characterize these objects.  A revised scheme, preferably based
on red optical and/or $J$-band spectral features, should be
matched closely to established L field dwarf schemes
(e.g., Kirkpatrick et al.\ 1999) to enable the straightforward assessment of
metallicity effects.  However, the definition of a rigorous classification scheme, as well as
more general studies of the L subdwarf population as a whole, require a systematic
search for these objects.  This may best be accomplished by a wide-field NIR proper motion survey,
in analogy to the successful photographic plate surveys of the past.

\acknowledgments

The author thanks telescope operator P.\ Sears
and instrument specialist J.\ Rayner for their assistance during the IRTF observations,
M.\ Cushing for pointing out the TiH band identification and for useful discussions,
G.\ Bj{\o}rn for logistical support during the preparation of
the manuscript, and the anonymous referee for her/his helpful critique of the original manuscript.
Support for this work was provided by NASA through the SIRTF Fellowship
Program.
POSS-II images were obtained from the Digitized Sky Survey
image server maintained by the Canadian Astronomy Data Centre,
which is operated by the Herzberg Institute of Astrophysics,
National Research Council of Canada. The Digitized Sky Survey was produced at the
Space Telescope Science Institute under U.S. Government grant NAG W-2166.
This publication makes use of data from the Two
Micron All Sky Survey, which is a joint project of the University
of Massachusetts and the Infrared Processing and Analysis Center, and
funded by the National Aeronautics and Space Administration and
the National Science Foundation.
2MASS data were obtained from
the NASA/IPAC Infrared Science Archive, which is operated by the
Jet Propulsion Laboratory, California Institute of Technology,
under contract with the National Aeronautics and Space
Administration.
The author wishes to extend special thanks to those of Hawaiian ancestry
on whose sacred mountain we are privileged to be guests.


Facilities: \facility{IRTF}{:SpeX}, \facility{Palomar}{:IRCam}.

\begin{deluxetable}{lc}
\tabletypesize{\small}
\tablecaption{Observational Properties of the L Subdwarf 2MASS 1626+3925.}
\tablewidth{0pt}
\tablehead{
\colhead{Parameter} &
\colhead{Value} \\
}
\startdata
${\alpha}$\tablenotemark{a} & 16$^h$ 26$^m$ 20$\farcs$34 \\
${\delta}$\tablenotemark{a} & +39$\degr$ 25$^m$ 19$\farcs$0 \\
$R$\tablenotemark{b} & 19.84 \\
$I_N$\tablenotemark{b} & 16.68 \\
2MASS $J$ & 14.44$\pm$0.03 \\
2MASS $J-H$ & $-$0.10$\pm$0.06 \\
2MASS $H-K_s$ & 0.07$\pm$0.09 \\
$\mu$ & 1$\farcs$27$\pm$0$\farcs$03 yr$^{-1}$ \\
$\theta$ & $282{\fdg}6{\pm}1{\fdg}4$ \\
$d$\tablenotemark{c} & $\sim$15-35 pc \\
$V_{tan}$\tablenotemark{c} & $\sim$90-210 km s$^{-1}$ \\
$V_{rad}$ & $-260{\pm}35$ km s$^{-1}$ \\
\enddata
\tablenotetext{a}{2MASS coordinates, equinox J2000 and epoch 29 April 1998 (UT).}
\tablenotetext{b}{Photographic $R$ (IIIaF) and $I_N$ (IV-N) magnitudes from USNO-B1.0 \citep{mon03}.}
\tablenotetext{c}{Assuming $M_J \approx 11.5-13.5$; see $\S$ 3.}
\end{deluxetable}

\begin{figure}
\centering
\epsscale{.90}
\plotone{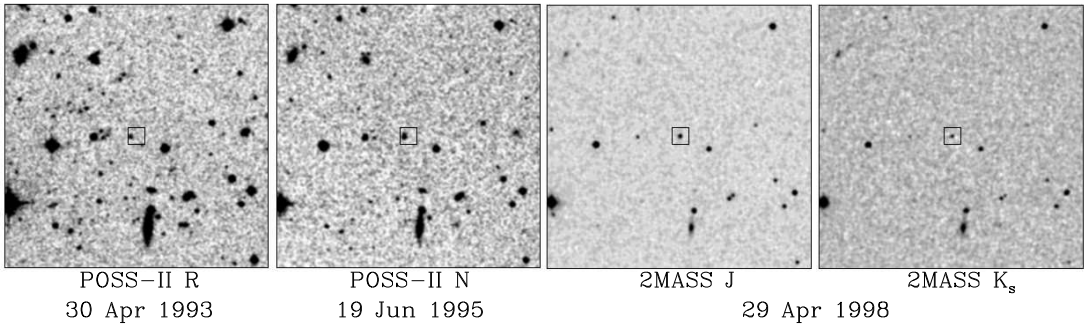}
\caption{POSS-II ($R$ and $I_N$) and 2MASS ($J$ and $K_s$)
images of the 2MASS 1626+3956 field.
Images are 5$\arcmin$ on a side and oriented with North up and East to the left.
A 10$\arcsec$ diameter box is centered at the 2MASS position of the
source.  The motion of 2MASS 1626+3956 is clearly visible over the 5 year
period shown here.
\label{fig1}}
\end{figure}

\begin{figure}
\epsscale{.70}
\plotone{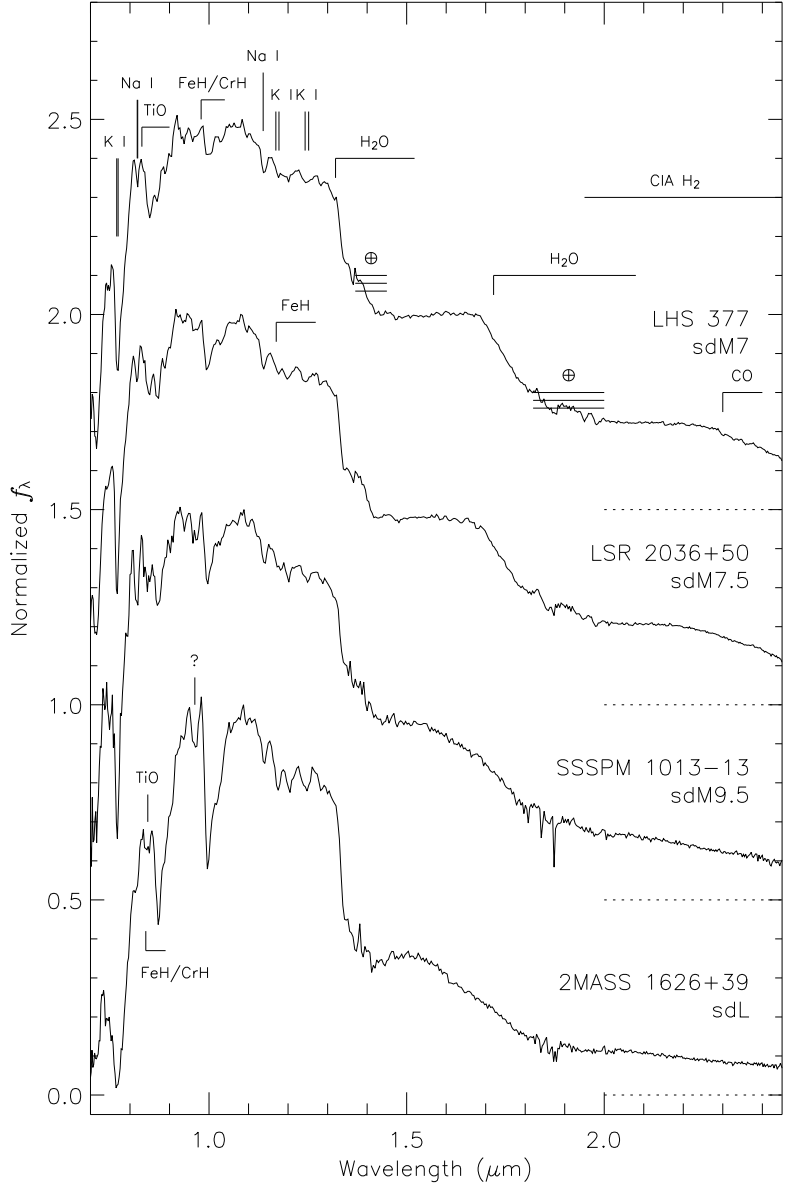}
\caption{SpeX prism spectra of the late M subdwarfs LHS 377 (sdM7),
LSR 2036+5059 (sdM7.5), and SSSPM 1013$-$1356 (sdM9.5) compared
to that of 2MASS 1626+3956.  Spectra are normalized at their flux peaks
and offset by constants (dotted lines). Key spectral features are labelled,
with the exception of the as yet unidentified band centered at 0.96 $\micron$ (B03).
Regions of high telluric absorption are indicated by the $\earth$ symbol.
\label{fig2}}
\end{figure}

\begin{figure}
\epsscale{1.0}
\plotone{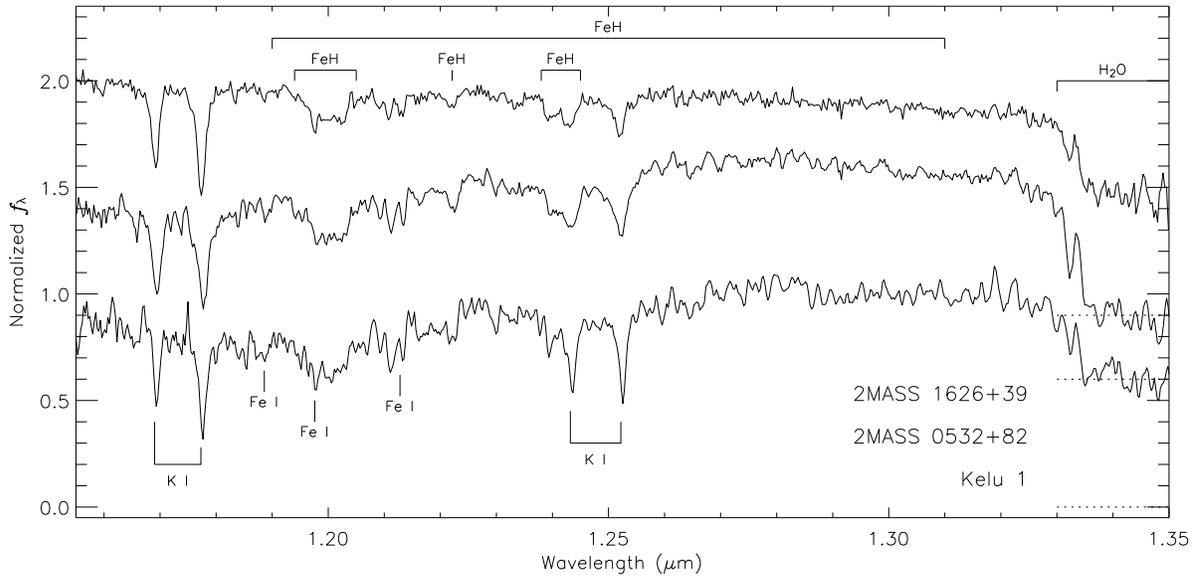}
\caption{Moderate resolution spectra of 2MASS 1626+3956 in the 1.15-1.35 $\micron$
region compared to NIRSPEC spectra of the L subdwarf 2MASS 0532+8246 (B03) and
the rapidly rotating L2 field dwarf Kelu 1 \citep{mcl03}.
Spectra are normalized at 1.27 $\micron$ and offset by constants (dotted lines).
Strong features arising from FeH, H$_2$O, and \ion{K}{1} are noted,
as are weaker lines of \ion{Fe}{1}.  Note the substantially broadened \ion{K}{1} line
wings in the L subdwarf spectra.
\label{fig3}}
\end{figure}


\begin{thebibliography}{}

\bibitem[Andersson et al.(2003)]{and03}Andersson, N., Balfour, W.\ J., Bernath, P.\ F.,
Lindgren, B., \& Ram, R.\ S. 2003, JChPh, 118, 3543

\bibitem[Bailer-Jones(2004)]{bai04} Bailer-Jones, C.\ A.\ L. 2004, \aap, 419, 703

\bibitem[Basri et al.(2000)]{bas00}Basri, G., Mohanty, S., Allard, F.,
Hauschildt, P.\ H., Delfosse, X., Mart{\'{\i}}n, E.\ L., Forveille, T.,
\& Goldman, B. 2000, \apj, 538, 363

\bibitem[Borysow, J{\o}rgensen, \& Zheng(1997)]{bor97}Borysow, A., J{\o}rgensen,
U.\ G., \& Zheng, C. 1997, \aap, 324, 185

\bibitem[Burgasser et al.(2003a)]{me03f}Burgasser, A.\ J., Kirkpatrick,
J.\ D., Burrows, A., Liebert, J., Reid, I.\ N., Gizis, J.\ E., McGovern, M.\ R.,
Prato, L., \& McLean, I.\ S. 2003a, \apj, 592, 1186 (B03)

\bibitem[Burgasser et al.(2003b)]{me03a}Burgasser, A.\ J., Kirkpatrick,
J.\ D., McElwain, M.\ W., Cutri, R.\ M., Burgasser, A.\ J., \& Skrutskie, M.\ F.
2003b, \aj, 125, 850


\bibitem[Burgasser et al.(2004)]{me04a}Burgasser, A.\ J., McElwain, M.\ W.,
Kirkpatrick, J.\ D., Cruz, K.\ L., Tinney, C.\ G., \& Reid, I.\ N. 2004, \aj, 127, 2856

\bibitem[Burgasser et al.(2002)]{me02a} Burgasser, A.\ J., et al. 2002, \apj, 564, 421

\bibitem[Burrows et al.(2002)]{bur02} Burrows, A., Burgasser, A.\ J., Kirkpatrick, J.\ D.,
Liebert, J., Milsom, J.\ A., Sudarsky, D., \& Hubeny, I. 2002, \apj, 573, 394

\bibitem[Burrows et al.(2001)]{bur01}Burrows, A., Hubbard, W.\ B., Lunine,
J.\ I., \& Liebert, J. 2001, Rev.\ of Modern Physics, 73, 719

\bibitem[Burrows, Marley, \& Sharp(2000)]{bur00}Burrows, A., Marley, M.\ S.,
\& Sharp, C.\ M. 2000, \apj, 531, 438

\bibitem[Cushing et al.(2003)]{cus03}Cushing, M.\ C., Rayner, J.\ T.,
Davis, S.\ P., \& Vacca, W.\ D. 2003, \apj, 582, 1066

\bibitem[Cushing, Vacca, \& Rayner(2004)]{cus04} Cushing, M.\ C., Vacca, W.\ D., \& Rayner, J.\ T.
2004, PASP, 116, 362

\bibitem[Cutri et al.(2003)]{cut03}Cutri, R.\ M., et al. 2003,
\url{http://www.ipac.caltech.edu/2mass/releases/allsky/doc/explsup.html}

\bibitem[Digby et al.(2003)]{dig03}Digby, A.\ P., Hambly, N.\ C., Cooke, J.\ A., Reid, I.\ N.,
\& Cannon, R.\ D. 2003, /mnras, 344, 583

\bibitem[Epchtein et al.(1997)]{epc97}Epchtein, N., et al. 1997,
The Messenger, 87, 27

\bibitem[Gizis(1997)]{giz97}Gizis, J.\ E. 1997, \aj, 113, 806

\bibitem[Golimowski et al.(2004)]{gol04} Golimowski, D.\ A., et al. 2004, \aj, 127, 3516

\bibitem[Hansen(1998)]{han98}Hansen, B.\ M.\ S. 1998, Nature, 394, 860

\bibitem[Kirkpatrick et al.(2000)]{kir00}Kirkpatrick, J.\ D., Reid, I.\ N.,
Liebert, J., Gizis, J.\ E., Burgasser, A.\ J., Monet, D.\ G., Dahn, C.\ C.,
Nelson, B., \& Williams, R.\ J. 2000, \aj, 120, 447

\bibitem[Kirkpatrick et al.(1999)]{kir99}Kirkpatrick, J.\ D., et al. 1999,
\apj, 519, 802

\bibitem[Leggett et al.(2000)]{leg00}Leggett, S.\ K., Allard, F., Dahn, C.,
Hauschildt, P.\ H., Kerr, T.\ H., \& Rayner, J. 2000, \apj, 535, 965

\bibitem[Lepine, Rich, \& Shara(2003)]{lep03a} Lepine, S.\
Rich, R.\ M., \& Shara, M.\ M. 2003, \aj, 125, 1598

\bibitem[Lepine, Rich, \& Shara(2003)]{lep03b} ---. 2003b, \apj, 591, L49

\bibitem[Lepine, Shara, \& Rich(2002)]{lep02} Lepine, S.\ Shara, M.\ M., \&
Rich, R.\ M. 2002, \aj, 124, 1190

\bibitem[Lodders(1999)]{lod99}Lodders, K. 1999, \apj, 519, 793

\bibitem[Luyten(1979)]{luy79a}Luyten, W.\ J. 1979, LHS Catalogue: A Catalogue of Stars with Proper
Motions Exceeding 0$\farcs$5 Annually (Minneapolis: Univ.\ Minn.\ Press)

\bibitem[Mart{\'{\i}}n et al.(1999)]{mrt99}Mart{\'{\i}}n, E.\ L., Delfosee, X.,
Basri, G., Goldman, B., Forveille, T., \& Zapatero Osorio, M.\ R. 1999, \aj, 118, 2466

\bibitem[McLean et al.(2003)]{mcl03}McLean, I.\ S., McGovern, M.\ R., Burgasser, A.\ J.,
Kirkpatrick, J.\ D., Prato, L., \& Kim, S. 2003, \apj, 596, 561

\bibitem[Monet et al.(1998)]{mon98}Monet, D.\ G., et al. 1998,
USNO-A2.0 Catalog (Flagstaff: USNO)

\bibitem[Monet et al.(2003)]{mon03} ---. 2003, \aj, 125, 984 (USNO-B1.0 Catalog)

\bibitem[Murphy et al.(1995)]{mur95}Murphy, D.\ C., Persson, S.\ E.,
Pahre, M.\ A., Sivaramakrishnan, A., \& Djorgovski, S. G. 1995, \pasp, 107, 1234

\bibitem[Rayner et al.(2003)]{ray03} Rayner, J.\ T., Toomey, D.\ W., Onaka, P.\ M., Denault, A.\ J.,
Stahlberger, W.\ E., Vacca, W.\ D., Cushing, M.\ C., \& Wang, S. 2003, PASP, 155, 362

\bibitem[Ruiz, Leggett, \& Allard(1997)]{rui97}Ruiz, M.\ T., Leggett,
S.\ K., \& Allard, F. 1997, \apj, 491, L107

\bibitem[Reid et al.(2002)]{rei02}Reid, I.\ N., Kirkpatrick,
J.\ D., Liebert, J., Gizis, J.\ E., Dahn, C.\ C., \&
Monet, D.\ G. 2002, \aj, 124, 519

\bibitem[Reid et al.(1991)]{rei91}Reid, I.\ N., et al. 1991, \pasp, 103, 661

\bibitem[Saumon et al.(1994)]{sau94}Saumon, D., Bergeron, P., Lunine, J.\ I.,
Hubbard, W.\ B., \& Burrows, A. 1994, \apj, 424, 333

\bibitem[Scholz et al.(2004)]{sch04} Scholtz, R.-D., Lehmann, I., Matute, I.,
\& Zinnecker, H. 2004, \aap, in press (astro-ph/0406457)

\bibitem[Vrba et al.(2004)]{vrb04}Vrba, F.\ J., et al. 2004, \aj, 127, 2948

\bibitem[York et al.(2000)]{yor00} York, D.\ G., et al. 2000, AJ,
120, 1579

\end{thebibliography}
\end{document}